\documentclass[aps,pra,reprint,showpacs,superscriptaddress,twocolumn,10pt]{revtex4-1}%
\usepackage{bm}
\usepackage{amsmath}
\usepackage{amssymb}
\usepackage{amsthm}
\usepackage{amsfonts}
\usepackage{mathrsfs}
\usepackage{graphicx}
\usepackage{color}%
\usepackage{mathtools}
\usepackage{booktabs}
\usepackage{mleftright}
\usepackage{float}
\usepackage[colorlinks,linkcolor=blue,citecolor=blue,urlcolor=blue,hyperindex,bookmarks=false,pdfstartview=FitH]{hyperref}
\usepackage[utf8]{inputenc}
\date{\today}
\begin{document}
\bibliographystyle{plainnat}

\title{Diversity of Superradiant Phase Transitions in the Bose-Fermi System under Tight-Binding Model in the Weak-Coupling Regime}

\author{Xing Su}
\affiliation{School of Physics and Information Technology, Shaanxi Normal University, Xi'an 710119, People Republic of China}

\author{Jian-Jian Cheng}

\affiliation{School of Science, Xi’an University of Posts and Telecommunications, Xi’an 710121, China}

\author{Lin Zhang}
\email{zhanglincn@snnu.edu.cn}

\affiliation{School of Physics and Information Technology, Shaanxi Normal University, Xi'an 710119, People Republic of China}

\begin{abstract}
We present a comprehensive analysis of the dynamic diversity associated with superradiant phase transitions within a one-dimensional tight-binding electronic chain that is intrinsically coupled to a single-mode optical cavity. By employing the quantized electromagnetic vector potential through the Peierls substitution, the gauge-invariant coupled Bose-Fermi system facilitates momentum-dependent superradiant transitions and effectively avoids the second-order spurious phase transitions typically observed in Dicke-like models. The quantum phase transitions in this system are characterized by stable dynamics, including the displacement and squeezing of the cavity mode and the redistribution of electronic momentum in the solid chain. Distinct from multimode cavity QED systems with atomic gases, this single-mode optical configuration unveils a range of nonlinear phenomena, including multistability and diversity of spontaneous symmetry breaking. The setup allows for precise manipulation of superradiant phases in the weak coupling regime, effectively mitigating the adverse effects of quantum fluctuation divergences. The diverse attributes of these quantum phase transitions enhance our understanding of tunable quantum solid devices and underscore their potential applications in quantum information processing and metrology.
\end{abstract}

\maketitle

\section{Introduction}

The manipulation of macroscopic quantum states through phase transitions is a cornerstone of modern science and has driven rapid advancements in quantum computing, information processing, and quantum communication~\cite{Ritsch, Matthias Vojta}. In recent years, superradiant phase transitions, as a collective emission phenomenon, have attracted considerable attention across a variety of systems, including two-dimensional electron systems, ultracold atoms as well as multiqubit systems~\cite{Wu2023Signatures,CollapseBEC2016,Zhang,Zheng}. By integrating matter into high-quality optical cavities, cavity quantum electrodynamics (cQED) significantly enhances the interaction between light and matter, creating an ideal platform for observing and controlling superradiant phase transitions~\cite{CavityMediatedTuning2023,EnhancingCQED2018,LightMatterWaveguide2023}. A deeper understanding and precise manipulation of these transitions not only contribute to uncovering fundamental physical principles but also provide a foundation for the development of advanced quantum devices and materials. As a result, the study of cavity-mediated superradiant phase transitions holds significant theoretical importance and offers promising potential for practical applications.

The conventional framework for describing light-matter interactions and phase transitions is provided by the Dicke model, which examines a collection of two-level atoms coupled to a single-mode light field for enhanced emission~\cite{NonreciprocalDicke2023,RevisitingDickeTransitions2022}. However, advancements in experimental techniques have opened new avenues for exploring light-matter interactions in more complex systems, such as those involving multimode light fields or systems with strong electronic correlations~\cite{RojasRojas2024Analytic,DissipativeDicke2018}. These phase transitions in tightly bound systems have emerged as a promising platform for engineering solid-state quantum devices and enabling on-chip quantum technology~\cite{QuantumFloquet2023}. The tight-binding model offers the advantage of exact solvability, which helps circumvent spurious superradiant phase transitions caused by truncation approximation~\cite{SolvableModel2023} or previously known as the no-go theorem~\cite{Rza}. Additionally, it facilitates momentum conservation and enables explicit rigorous analysis of the energy density in electronic states~\cite{OpticalConductivity2021}.

In this work, we introduce a theoretical framework for understanding and designing quantum phase transitions in light-matter interactions within a hybrid Bose-Fermi system. Our model focuses on a one-dimensional tight-binding electronic chain coupled to a single spatially uniform cavity mode~\cite{Li2020}. Through this coupling and under the condition of a deliberately engineered momentum distribution of the electronic chain, the system achieves multiphoton beam emission~\cite{CavitySuperradiance2023}. Compared to earlier studies, this model distinguishes itself by employing a distinct physical mechanism—specifically, modifying the electronic distribution to break the symmetry of the light field and incorporating the nonlinear characteristics into the superradiant phenomena. This method broadens the experimental conditions required for achieving multiphoton beam emission, making it applicable to a wider range of systems~\cite{He2021}.

The Bose-Fermi system addresses limitations observed in the traditional Dicke model, where improper treatment of light-matter coupling can lead to false indicators of superradiant phase transitions~\cite{Wang1973,Lundgren2020}. Through the implementation of the quantized Peierls substitution, the tight-binding model avoids such spurious signals~\cite{Do2022}. Furthermore, the superradiant phase transition in this system occurs under weak interactions, requiring a lower critical coupling strength. This framework enables the exploration of superradiant phenomena in solid-state light-matter systems with enhanced nonlinear features~\cite{FirstOrderSuperradiant}. By incorporating the nonlinear aspects of phase transitions and leveraging the tight-binding framework, the study offers additional insights into the feature and control of quantum phase-transition phenomena in hybrid solid systems.

\section{Model and Analysis}

\begin{figure}[thpb]
		\centering
		\includegraphics[width=0.45\textwidth]{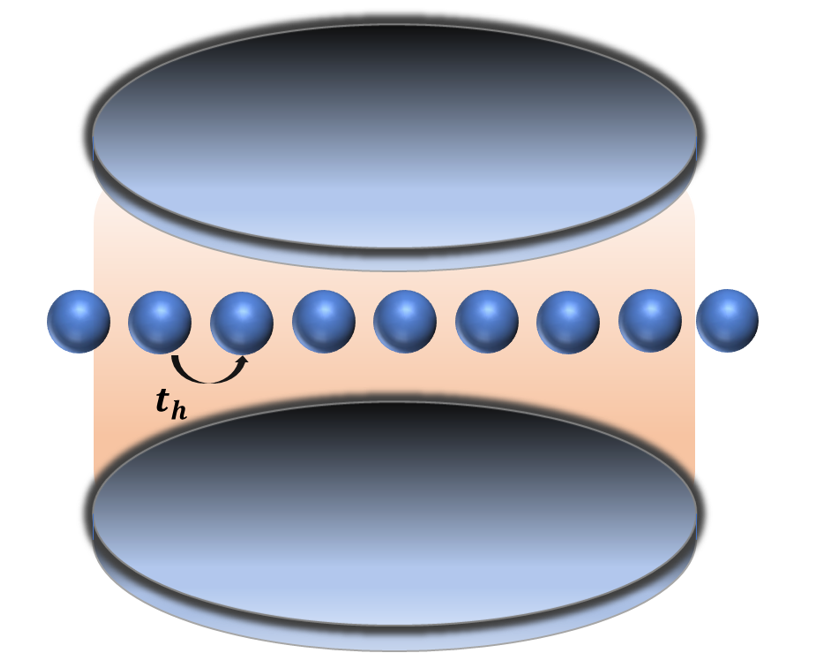} %
		\caption{ Illustration of the system model: A one dimensional tight-binding chain with nearest neighbor hopping $t_{h}$ is coupled to the first transmittance resonance of a cavity at $w_{0}$.}
		\label{fig1}
	\end{figure}

We investigate a model based on a one-dimensional tight-binding chain coupled to a single spatially uniform cavity mode as shown in Fig.~\ref{fig1}. The system consists of a non-interacting tight-binding chain with nearest-neighbor hopping $t_h$, which is coupled to the first transmission resonance mode of a Fabry-Pérot (FP) optical cavity.
In this system, we focus primarily on the continuous mode inside the cavity around frequency of $\omega_0$, assuming these modes have zero wave vector along the direction of the chain.  This setup is equivalent to the dipole approximation, where all off-resonant modes that couple weakly to the matter degrees of freedom are neglected due to strong suppression caused by the cavity geometry and boundary conditions. All the near-resonant modes are degenerately treated with the same frequency $\omega_0$, and assume that their coupling strength to the tight-binding chain, $g(\omega_0)$, is uniform. This can be represented by a box function centered at the frequency $\omega_0$ with a width of $\Delta \omega$. After simplifying the model, it can be reduced to a spatially uniform single-mode coupling system, and the corresponding Hamiltonian is expressed as \cite{QuantumFloquet2023}:
\begin{equation}
\label{H0}
\hat{H}= \omega_{0}\left(\hat{a}^{\dagger} \hat{a}+\frac{1}{2}\right) - \sum_{j = 1}^{L}\left[t_{h} e^{-i \frac{g}{\sqrt{L}}\left(\hat{a}^{\dagger}+\hat{a}\right)} \hat{c}_{j+1}^{\dagger} \hat{c}_{j} + \mathrm{H.c.}\right].
\end{equation}
where $\hat{a}^{\dagger}$ represents the bosonic creation operator for photons in the cavity mode, and $\hat{c}_{j}^{\dagger}$ represents the fermionic (electron) creation operator at site $j$ of the lattice. The parameter $t_{h}$ denotes the nearest-neighbor hopping energy of electrons, and ``H.c." stands for the Hermitian conjugate of the preceding term. The system is described in atomic units where $e = \hbar = c = 1$, and $L$ is the number of lattice sites. In the following calculations of this article, assume $t_{h}=1$ and $\omega_0=1$ by default.
		
To account for the light-matter coupling, we employ the Peierls substitution \cite{GaugeFixing2020}. In the presence of a quantized electromagnetic field, the hopping term between neighboring sites is modified by the gauge potential $\hat{A}$. This quantized vector potential is given by one quadrature of the cavity field:
\begin{equation}
\hat{A} = \frac{g}{\sqrt{L}}\left(\hat{a}^{\dag } + \hat{a}\right),
\end{equation}
where $\hat{A}$ represents the quantized electromagnetic vector potential, while $g$ denotes the coupling strength between light and matter, determined by factors such as the cavity geometry and the polarization of the atomic material. In this work, $g(\omega_0)\equiv g$ is modeled as a constant.

To simplify the calculations, we transform the Hamiltonian into quasi-momentum space. The Hamiltonian in momentum space takes the diagonal form as \cite{QuantumFloquet2023}	
\begin{equation}
\hat{H} = \omega_{0}\left(\hat{a}^{\dagger} \hat{a}+\frac{1}{2}\right) - 2 t_{h} \sum_{k} \cos (\hat{A} + k) \hat{c}_{k}^{\dagger} \hat{c}_{k}. \label{Hk}
\end{equation}

\begin{figure}[htpb]
	\centering
	\includegraphics[width=0.5\textwidth]{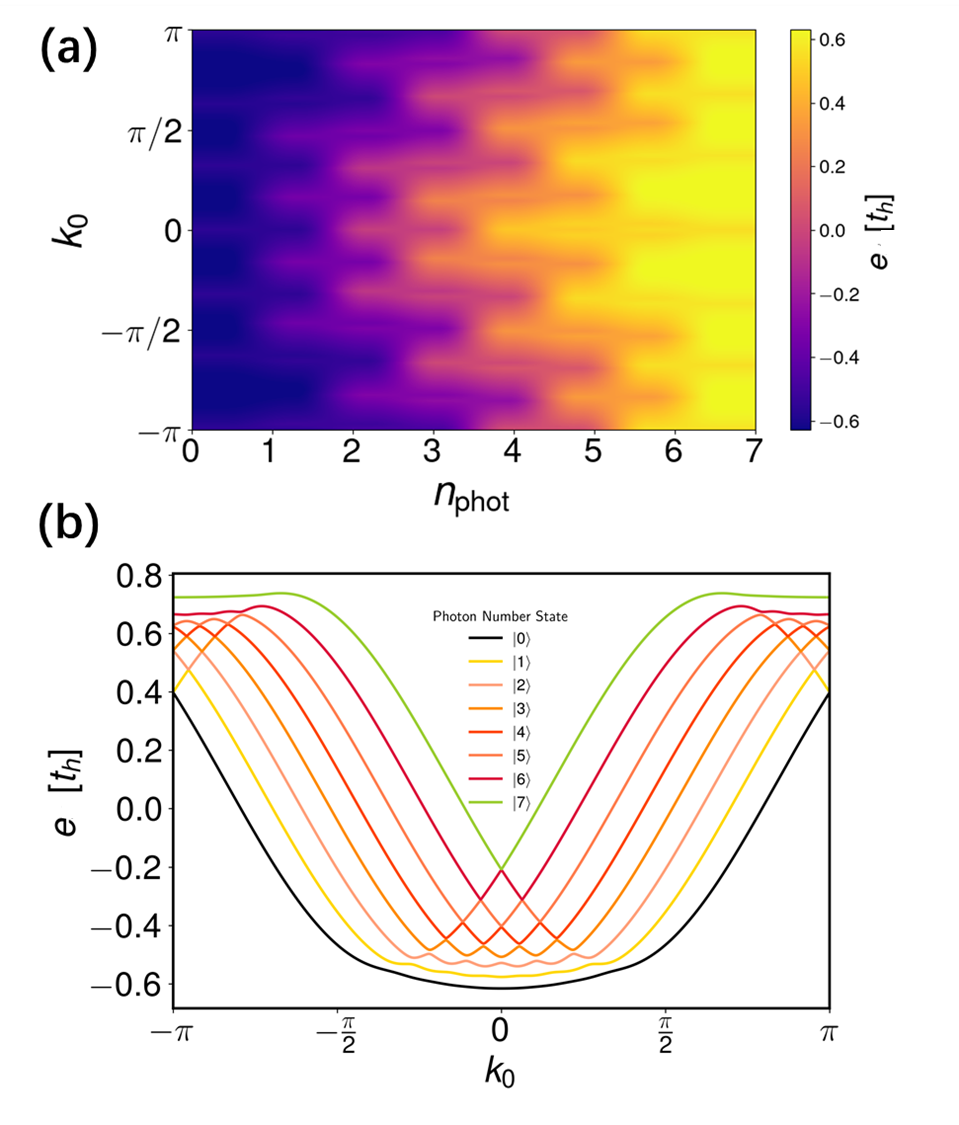}
	\caption{(a) It shows the energy density~$e =\frac{E}{L}$ changing with the electronic state of $k_0$ and the cavity state of different intensity with parameters $L=15$, and $g= 3$.
	(b) Energy density plain band for different photon number states of the cavity field (colored lines).
	 The electronic part of the wavefunction $\left | \psi  \right \rangle _{f} $ is chosen to occupy a single connected quasi-momentum region with parameters $L=60$, and $g= 2$.}
	\label{fig2}
\end{figure}

Here, $k$ denotes the quasi-momentum, and $\hat{c}_{k}^{\dagger}$ and $\hat{c}_{k}$ are the creation and annihilation operators for electrons in the quasi-momentum state $k$, respectively. The energy band dispersion relation $E_{d}(k) = t_{h}\cos(\hat{A} + k)$, shifted by the field vector potential $\hat{A}$, results in a flat energy band of ground state, as shown in Fig.\ref{fig2}, which facilitates a uniform ground-state population distribution for the electronic state. In this configurations, the system's energy density varies with Fermi center $k_{0}$ and $\hat{A}$ to form segments of flat band in the bottom of the energy spectrum under different vacuum states with half-filled population distributions $n(k)$ (see Fig.\ref{fig3}).
This feature helps to prevent misleading phase transition signals, maintains a stable energy reference, and simplifies band control of the system. Fig.\ref{fig2}(a) shows the energy density band of Eq.(\ref{Hk}) changing with the electronic state of $k_0$ and the cavity state of different intensity, which is represented by the photon number state $n_{phot}$. Fig.~\ref{fig2}(b) is a section figure of Fig.\ref{fig2}(a) for fixed intensity of cavity field. We can see a clear flat energy band versus $k_0$ by the energy band shift of the chain dispersion relation of$E_{d}(k)=t_{h}\cos(k)$ introduced by the interaction between electronic chain with the cavity field.

In the $k$-space, the excitation (occupation) number of the electronic states in the band, $\hat{\rho}_{k} =  \hat{c}_{k}^{\dagger} \hat{c}_{k}$, is conserved since $[\hat{\rho}_{k}, \hat{H}] = 0$. This indicates a profile preservation of the electronic wavepacket during the election-photon interactions. However, the photon number in the cavity, $\hat{a}^{\dagger} \hat{a}$, is not conserved. This non-conservation arises due to the non-commutative relationship between the coupling term $\cos (\hat{A}+k)$ and the photon number operator, given by the other field quadrature of
\begin{equation}
\left[\hat{a}^{\dagger} \hat{a}, \hat{A}\right] = \frac{g}{\sqrt{L}}\left(\hat{a}^{\dagger} - \hat{a}\right).
\end{equation}	
The non-commutation relationship demonstrates that the photon number (or the intensity of the cavity field) exhibits fluctuations as a result of electron-photon interactions. These fluctuations drive dynamical processes that are essential for the emergence of superradiant phase transitions within the cavity field. Furthermore, the band phase shifts induced by the cavity mode can self-consistently alter the synchronized band emission of the electronic chain. This modification facilitates the decoupling of the light field states from the collective ground electronic mode, which is attributed to the quasi-momentum conservation inherent in the electron-photon scattering process \cite{QuantumFloquet2023}. As a consequence, the eigenstates of the system's Hamiltonian, $|\Psi\rangle$, can be expressed as a direct product of the light field states and the electronic chain states:
\begin{equation}
\hat{H} \left\vert \Psi \right\rangle = E \left\vert \Psi \right\rangle, \quad \left\vert \Psi \right\rangle = \left\vert \phi \right\rangle_{b} \otimes \left\vert \psi \right\rangle_{f},
\end{equation}
where $\left\vert \phi \right\rangle_{b}$ is the bosonic wavefunction of the photon field in the cavity, and $\left\vert \psi \right\rangle_{f}$ is the fermionic wavefunction of the electron chain.
The decomposition indicates the complex interactions between light and electron chain accounting for the phase coupling through the Peierls substitution still enables a decoupling ground state in the thermodynamic limit due to an optimal electronic density distribution of the Fermi sea in the electron chain \cite{Rokaj}. However, the momentum tunability of the Fermi sea by the vector potential $\hat{A}$ endows this system with richer degrees of freedom, surpassing the capabilities of simpler models like the traditional Dicke model~\cite{FriedelOscillations}.

Overall, the flat energy bands induced by potential vector for cavity field provides easy going tunable platform for investigating light-matter interactions in low-dimensional systems, leading to a range of intriguing quantum phase transition phenomena. For instance, in cavity quantum electrodynamics (QED) systems, vacuum fluctuations can drive quantum phase transitions in the cavity field~\cite{CavityFerroelectric}. This coupling not only opens new avenues for understanding quantum phase transitions but also highlights the potential for designing quantum devices, such as enabling precise control over electron and photon states in superconductors and cold atoms in optical lattices. These findings offer fresh perspectives and tools for exploring light-induced quantum phase transitions and advancing applications in complex light-matter coupled systems~\cite{DrivenDissipativeControl}.

\section{Nonlinear Characteristics of the Phase Transition}

\begin{figure}[thpb]
	\centering
	\includegraphics[width=0.45\textwidth]{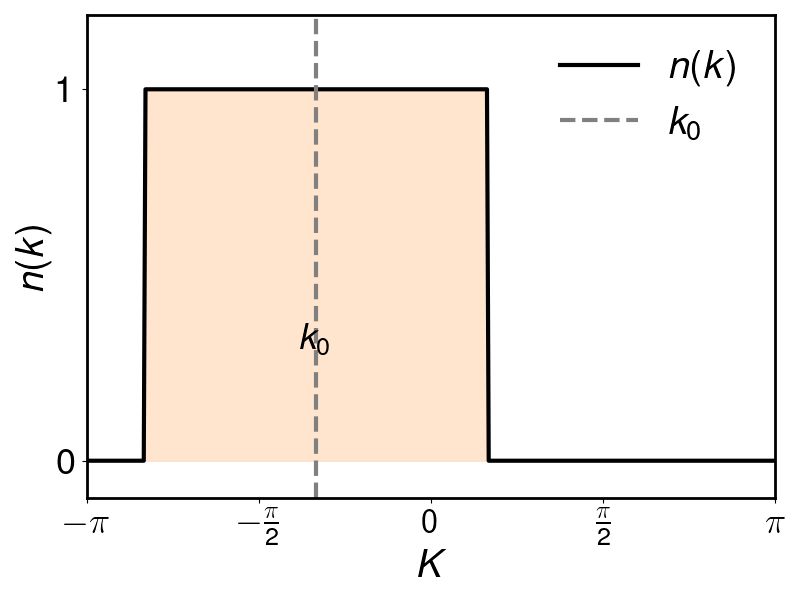} %
	\caption{Electron occupation number $n(k)$ as a function of momentum $k$ with half-filled Fermi sea surface centered at $k_{0}$.}
	\label{fig3}
\end{figure}

In order to analyze the phase-transition behavior of the cavity field coupling with an electronic chain for the thermodynamic limit, we consider an electronic ground state with a half-filled and continuously distributed momentum space around $k_0$ as shown in Fig.~\ref{fig3}.
Here, the ground-state electronic wavepacket can be thoroughly drifted by a bias voltage through an external electric field, effectively shifting the symmetric center $k_0$ of the Fermi sea in momentum space. Under these conditions, the Hamiltonian of the system can be integrated by using
\begin{eqnarray}
-t_{h}\frac{L}{\pi }\int_{k_{0}-\pi }^{k_{0}+\pi }\cos\left( k\right) n\left( k\right) \mathrm{d}k &=&-t_{h}\frac{L}{\pi }\int_{k_{0}-\frac{\pi }{2}}^{k_{0}+\frac{\pi }{2}}\cos \left( k\right) \mathrm{d}k \notag\\ &=&\frac{L}{\pi }\left[ -2t_{h}\cos \left( k_{0}\right) \right],
\end{eqnarray}
and
\begin{eqnarray}
	t_{h}\frac{L}{\pi }\int_{k_{0}-\pi }^{k_{0}+\pi }\sin
	\left( k\right) n\left( k\right) \mathrm{d}k &=&t_{h}\frac{L}{\pi }\int_{k_{0}-
		\frac{\pi }{2}}^{k_{0}+\frac{\pi }{2}}\sin \left( k\right) \mathrm{d}k\notag \\
	&=&\frac{L}{\pi }\left[ 2t_{h}\sin \left( k_{0}\right) \right],
\end{eqnarray}
which gives a tractable Hamiltonian from Eq.\,(\ref{Hk}) in case of a continuous half-full distribution of electrons as
\begin{align}
\hat{H}_g = \omega_{0}\left(\hat{a}^{\dagger}\hat{a} + \frac{1}{2}\right)-\frac{L}{\pi} 2 t_{h}\cos\left ( \hat{A} + k_{0} \right ) ,\label{ok}
\end{align}
where $k_0$ represents the symmetric center of the half-filled Fermi sea of the electronic chain.
From Eq.\,(\ref{ok}), we can see that the energy density of the system can be easily manipulated by momentum shift through an external electric field. The quantized electromagnetic vector potential $\hat{A} $ addresses the coupling between the electronic states and the cavity photons.
	
To analyze the system's nonlinear behavior, we employ the Heisenberg equations of motion. Here, we define the quadrature operators of the cavity field
\begin{align}
\hat{X} = \frac{1}{\sqrt{2}}\left(\hat{a}^{\dagger} + \hat{a}\right), \quad \hat{Y} =\frac{i}{\sqrt{2}}\left(\hat{a}^{\dagger} - \hat{a}\right),
\end{align}
which satisfy the commutation relation $[\hat{X}, \hat{Y}] = i$. In terms of these quadrature operators, the Hamiltonian for the entire system becomes
\begin{align}
\hat{H}_g = \frac{1}{2} \omega_{0} (\hat{X}^{2} + \hat{Y}^{2} ) - \frac{2 t_{h} L}{\pi} \cos \left(k_{0} + \frac{\sqrt{2} g}{\sqrt{L}} \hat{X}\right).
\end{align}	

	\begin{figure*}[htpb]
	\centering
	\includegraphics[width=1.\textwidth]{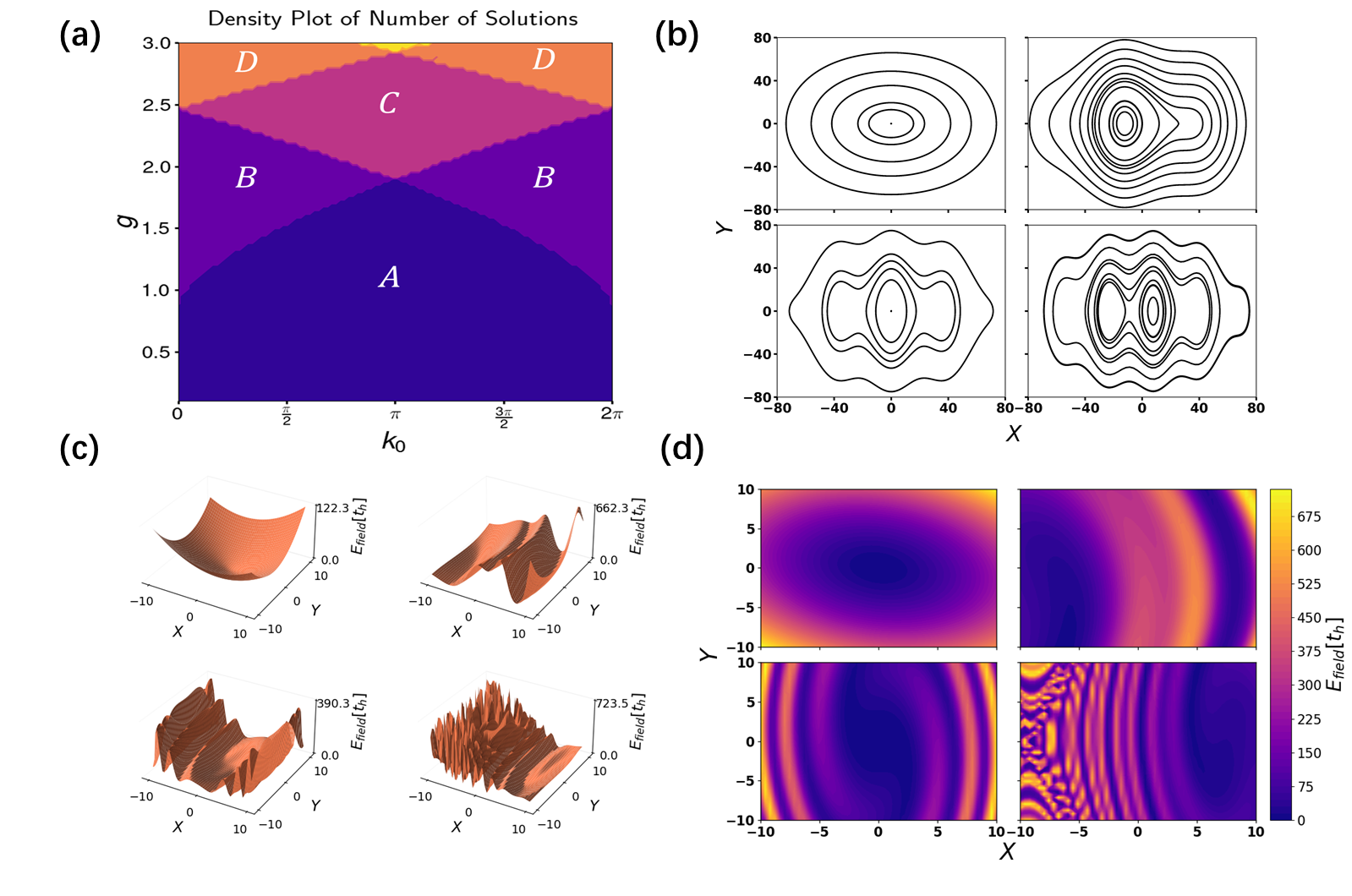} %
	\caption{Nonlinear analysis diagram with parameters $L = 510$.		
		(a) The phase diagram is divided into four regions A, B, C, and D based on the number of equilibrium steady-states of the cavity field, $1,3, 5$ and $7$.
		(b) From left to right and top to bottom, the corresponding dynamic trajectories in the phase space for A, B, C, and D cases are shown. The parameters for these regions are $g = 0.5, 1.5, 2.5, 3$ and $k_0 = 0, \pi/2, \pi, 3\pi/2$.		
		(c) The corresponding energy surface of the system. The $z$-axis represents the field energy $E_{field}$ and the parameters are the same as those in (b).		
		(d) The contour plot of the energy surface in part (c) with the cross section at the lowest energy point.}
	\label{fig4}
\end{figure*}

To reveal the dominant nature of the cavity field, we assume that the quantum fluctuations in quadratures $\hat{X}$ and $\hat{Y}$ are relative small to mean-value variations. We expand the quadrature operators around their mean displacements, $\hat{X}= \langle \hat{X} \rangle + \delta \hat{X}\equiv X+\delta \hat{X}, \hat{Y}= \langle \hat{Y} \rangle + \delta \hat{Y}\equiv Y+\delta \hat{Y}$, and substitute them back into the Heisenberg equations of motion. By neglecting second-order and higher-order fluctuation terms, we obtain a set of coupled nonlinear differential equations for the mean-field dynamics of quadrature displacements	
\begin{align}
\frac{d X}{d t} &= \omega_{0} Y,  \label{xt}\\
\frac{d Y}{d t} &= -\omega_{0} X + \frac{2 t_{h} g \sqrt{2 L}}{\pi} \sin \left(\frac{\sqrt{2} g}{\sqrt{L}} X - k_{0}\right), \label{yt}
\end{align}	
and the fluctuation equation of $\delta \hat{X}$
\begin{equation}
\frac{d^{2}}{dt^{2}}\delta \hat{X}+\omega^{2}\delta \hat{X}=0, \label{Dxt}
\end{equation}%
where the effective frequency $\omega$ for the quantum fluctuation of $X$ quadrature is
\begin{equation}
\omega =\sqrt{\omega^{2} _{0}-\frac{4 t_{h} g^{2}}{L\pi}\cos \left( \frac{\sqrt{2}g%
}{\sqrt{L}}X-k_0\right)}. \label{Omegat}
\end{equation}

Eq.\,(\ref{Dxt}) lays down a weak coupling condition on field fluctuations for continuous phase-transition analysis of the second order, that is
\begin{equation}
\label{deltx}
\frac{2g}{\omega_0}\leq \sqrt{\frac{\pi L}{t_h}},
\end{equation}
which maintain the stable classical dynamics with collapse and revival of the quantum fluctuations to validate the mean-value dynamics of the cavity field. This condition is broken due to the divergence of the quantum fluctuations at a critical point in the first-order phase transition in this system, and thus gives the ultrastrong coupling condition of the discontinuous phase transition from the normal phase to superradiant phase \cite{Hioe}. Obviously, for a large-size chain interacting with a light field, a weak coupling condition of Eq.(\ref{deltx}) with limited quantum fluctuations can be easily satisfied.

Mathematically, Eqs.(\ref{xt})-(\ref{yt}) follow the similar equations to the Van der Pol-Duffing (VDP-Duffing) oscillator, which is known to exhibit rich nonlinear features, such as multistability and chaotic dynamics \cite{RareAttractors2023,BifurcationChaos2023}.  To study the nonlinear properties of the system, we use the Jacobian criterion to construct the phase diagram of the system, as shown in Fig.\ref{fig4}(a), analyze the dynamical behavior of the cavity field under different phases, as depicted in Fig.\ref{fig4}(b), and examine the energy distribution presented in Fig.~\ref{fig4}(c). As indicated by the energy surfaces in Fig.~\ref{fig4}(c) and its cross-sectional view in Fig.~\ref{fig4}(d), the phase transitions are characterized by a shift from single-well to multi-well potential landscape. Moreover, at the point of minimum energy for the photon field, both $X$ and $Y$ components are non-zero, suggesting that the ground state may involve multi-photon generation. During this process, the photon field may switch between different stable states under the influence of external perturbations or noise. We need to be mindful of these perturbations, as they may disrupt the transitions.

From Eq.(\ref{ok}) and Eq.(\ref{yt}), we can easily find that the mean-value density energy of the system for a stable varying state of cavity field around equilibrium points, $X_{\ast}\sim \sqrt{\frac{L}{2}}(m\pi +k_0)/g $ with $ m\in \mathbb{Z} $, will be
\begin{equation}
E_g(t)=\langle\hat{H}_g\rangle/L=n\omega_{0}-\frac{2 t_{h}}{\pi} \cos\left[X(t) +k_{0} \right],  \label{Energyt}
\end{equation}
where $n=\langle \hat{a}\hat{a}^{\dagger}\rangle$ is the average photon number and the motion of $X(t)$ around equilibrium value $X_{\ast}$ is $X(t)=R_0 \sin \left[\sqrt{\omega_{0}(\omega_{0} -\frac{4g^2}{\pi})}\cdot t+\phi_0 \right]$ with $R_0$ being its varying amplitude. For a population distribution of electron states around $k_0$, the mean energy density can be
\begin{equation}
\bar{E}_g=\langle E_g(t)\rangle\approx n\omega_{0}-\frac{2t_{h}}{\pi}J_{0}(R_0)\cos(k_0),  \label{Et}
\end{equation}
where the mean energy density $n\omega_0$ of the cavity field is modified by the first order Bessel function $J_0(R_0)\cos(k_0)$ and can be verified by Fig.~\ref{fig2}(a). This result shows that the dynamic of quadrature $X(t)$ around its stable states can also lead energy band shifts according to Eq.\,(\ref{ok}) which strongly depends on the nonlinear characteristic of the phase transition for the cavity field.

\begin{figure}[htpb]
	\centering
	\includegraphics[width=0.5\textwidth]{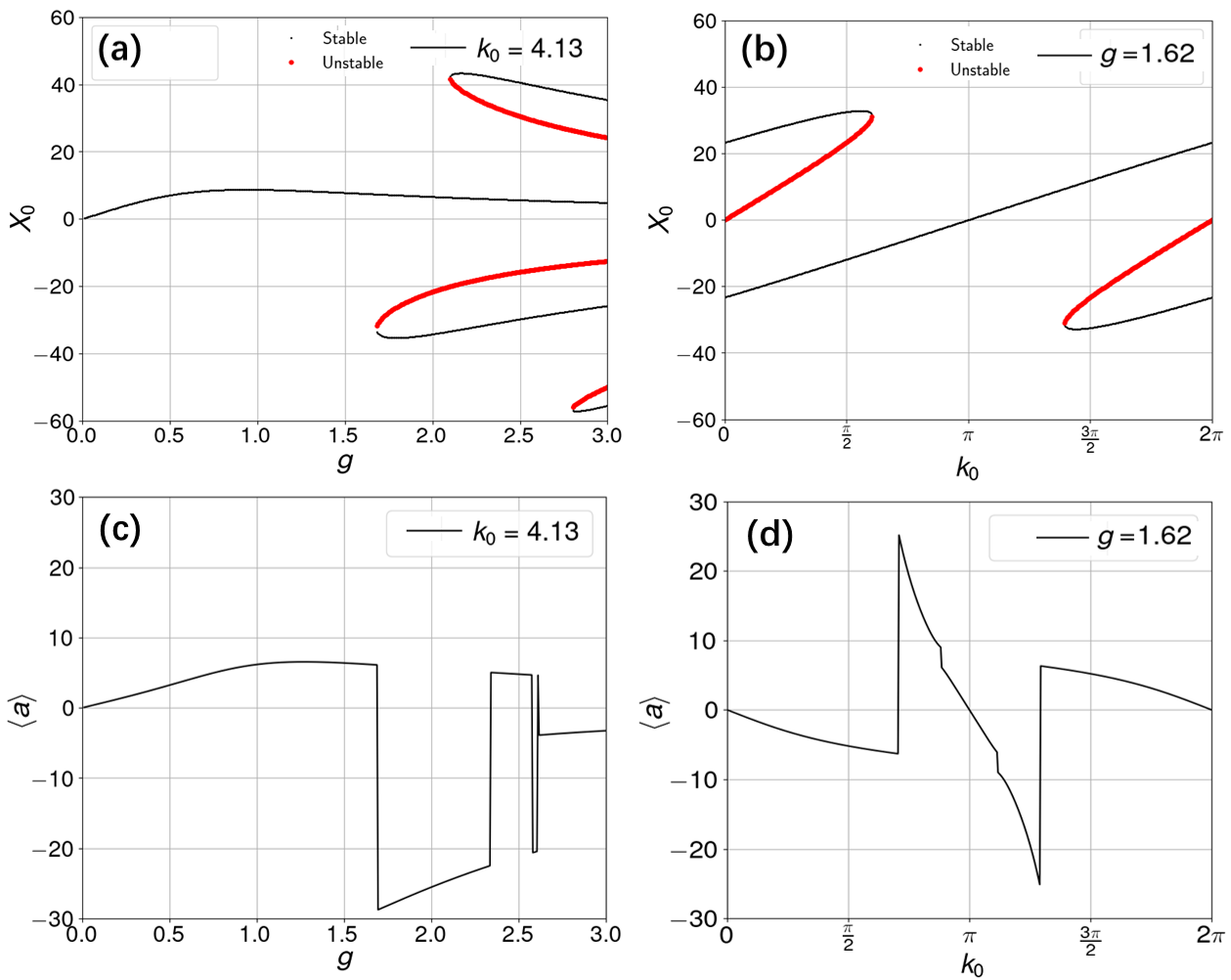}
	\caption{The bifurcations of the cavity mode $X(t)$ around steady states versus $g$ with fixed $k_0=4.13$ in (a) and versus $k_0$ with fixed $g=1.62$ in (b). The corresponding order parameters $\langle \hat{a}\rangle$ of (a) and (b) are (c) and (d) respectively, with chain length  $L=510$.}
	\label{fig5}
\end{figure}

Moreover, as the parameter $k_0$ varies, the whole system may exhibit spontaneous symmetry breaking. This phenomenon is of significant importance in nonlinear optics, particularly in applications such as mode selection and beam shaping. This result reveals that the ground state of the photon field in this model can achieve a non-trivial quantum coherent state, which has important physical implications for understanding the nonlinear dynamical behavior of light-matter interactions \cite{DissipativePhotonTransitions2019,GroundStateQuench2019}.

Now, by using the traditional semi-classical method in Dike models \cite{Wang1973}, we can directly demonstrate that the diversity of the dynamical transitions shown in Fig.\ref{fig4} represents a rich feature of the superradiant phase transition in this system. Unlike the traditional models where superradiance is often associated with uniform atomic ensembles, the introduction of a tight-binding chain and a spatially uniform cavity mode creates a new canonical partition function in the thermodynamic limit to govern the superradiant phase transition as
\begin{eqnarray}
Z_g(T) &=&\int \frac{d^2\alpha}{\pi}\langle\alpha | e^{-\beta \hat{H}_g}|\alpha\rangle \notag \\
&=&\int \frac{d^2\alpha}{\pi} e^{-\beta \omega_0 |\alpha|^2 }e^{\beta \frac{2Lt_h}{\pi}\cos\left[\frac{g}{\sqrt{L}}(\alpha+\alpha^{\ast})+k_0\right]},
\end{eqnarray}
where $|\alpha\rangle$ is the coherent state of the cavity field, $\beta=1/k_B T$ and $T$ is the temperature for the system ensemble. If we set $\alpha=x+iy$ and use the Laplace's method of integration for low temperature limit, we obtain
\begin{eqnarray*}
Z_{g}(T\rightarrow 0) &=&\frac{1}{\sqrt{\pi \beta \omega _{0}}}\int_{-\infty }^{+\infty
}e^{-\beta \varphi \left( x\right) }dx, \\
&=&\frac{1}{\beta \sqrt{\omega _{0}\varphi ^{\prime \prime }\left(x_{\ast
}\right) }}e^{-\beta \varphi \left( x_{\ast }\right) },
\end{eqnarray*}%
where the Landau potential function $\varphi \left( x\right) $ is%
\begin{equation}
\varphi \left( x\right) =\omega _{0}x^{2}-\frac{2Lt_{h}}{\pi }\cos \left( \frac{%
2g}{\sqrt{L}}x+k_{0}\right) , \label{landp}
\end{equation}%
and $x_{\ast }$\ is the equilibrium points at which $\varphi^{\prime}\left(x_{\ast }\right) =0$. Obviously, the Landau potential given by Eq.(\ref{landp}) is just the band diagram of the system near the band bottom as shown in Fig.\ref{fig2}(b). According to the Landau phase transition theory, the
critical points of the phase transition can be determined by the minimum
value of the potential function for%
\begin{equation*}
\varphi ^{\prime }\left(x_{\ast}\right) =2\omega _{0}x_{\ast}+\frac{4g\sqrt{L}t_{h}}{\pi }%
\sin \left( \frac{2g}{\sqrt{L}}x_{\ast}+k_{0}\right) =0,
\end{equation*}%
which leads to the explicit equation%
\begin{equation}
\label{xc}
x_{\ast}=-\frac{2g\sqrt{L}t_{h}}{\omega _{0}\pi }\sin \left( \frac{2g}{\sqrt{L}}%
x_{\ast} +k_{0}\right).
\end{equation}
The existence of solutions to Eq.\,(\ref{xc}) indicates a multiple phase transition behaviors produced at different critical points $x_{\ast}$ where the bifurcations of $X(t)$ occur, and the phase transitions can be easily controlled by both the electronic momentum distribution center $k_0$ and the light-matter coupling parameter $g$. In the Dicke model, the expectation value of the cavity field's optical field amplitude is commonly used as the order parameter\cite{DynamicalPhaseTransition2015}. Therefore, we define the expectation value of the optical field operator, $ \langle a \rangle$, as the system's order parameter. As shown in Fig.~\ref{fig5}(a)(b), the rich dynamical bifurcations of quadrature $X(t)$ determined by Eq.\,(\ref{xt}) and Eq.\,(\ref{yt}) versus $g$ and $k_0$ reveal the multiplicity of the phase transitions in the cavity field and can be easily verified by the corresponding order parameters $\langle \hat{a}\rangle$ shown in Fig.~\ref{fig5}(c)(d).  These  results suggest that the interplay between electronic and optical degrees of freedom in this model can lead to a more flexible and tunable dynamical behaviors maintained in different phase regimes, with the potential applications ranging from quantum metrology to the development of new light sources.

\section{Superradiant Phase Transitions}

\begin{figure}[htpb]
	\centering
	\includegraphics[width=0.5\textwidth]{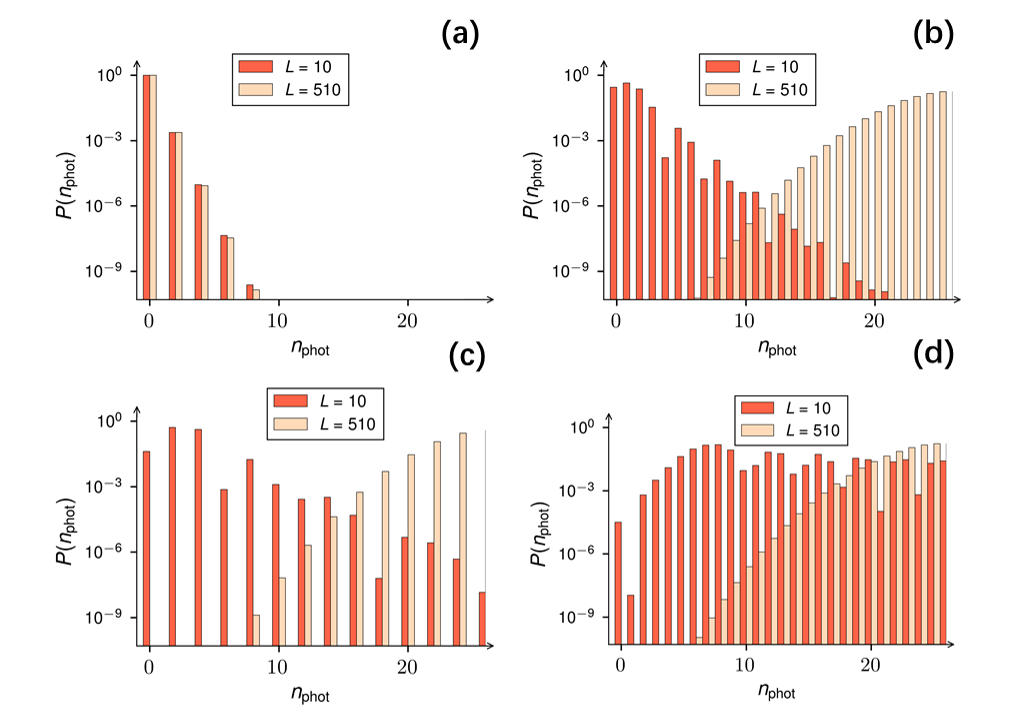} %
	\caption{ The logarithmic probability distribution of the photon number $P(n_{\text{phot}})$ corresponds to the ground state of the chain with lengths  $L=510$, $n_{max}=30$ (represents the maximum number of photons) and $ L=10$, where the parameters are the same as that in Fig.5(b)-(d).}
	\label{fig6}
\end{figure}

Based on the theoretical analysis above, we have clearly demonstrated the phase transition of the Fermi-Bose system for low temperature limit and the nonlinear diversity of the phase transitions. Now, in the case of a finite chain, we calculate the superradiant phase transition phenomenon of the cavity field by strictly solving the system's Schrödinger equation. As superradiance was initially proposed by Dicke in atomic gas system, describing the significant increase in photon number due to the collective interactions between a large ensemble of two-level atoms and the light field, some work has revealed that superradiant transitions can be suppressed by the nonlinear terms, such as $A^2$, of the light-matter interactions~\cite{QuantumFloquet2023}. In some cavity models, similar phenomena and problems have also been observed under different conditions \cite{SuperradiantBursts2024,MultimodeSuperradiance2024}.
To rule out the possibility of pseudo-superradiance in a finite system, we should investigate the ground state of the photon field by its photon number distribution, incorporating all orders of the Peierls transformation, to further explore superradiant phase transitions through strict solutions of systems which are shown in Fig.\ref{fig6}.

We define the photon ground state $|\phi_{GS}\rangle_b$ and introduce the photon number distribution $P(n_{phot})\equiv |\langle n_{phot}|\phi_{GS}\rangle|^2$ to indicate the superradiation of the cavity field through the probability of finding $n_{phot}$ photons of cavity ground state. For a symmetric distribution of the electron state with $k_0 = 0$, Fig.~\ref{fig6}(a) shows that $P(n_{phot})$ is only contributed by the even-numbered states, indicating that the photon distribution is incompatible with a coherent state but a squeezing state. This behavior, unaffected by the coupling strength $g$, suggests that no superradiant phase transition occurs in this case.
However, as $k_0$ increases to progressively break the symmetric electric distribution, the ground state photon population quantified by $n_{phot}\equiv \langle \hat{a}^{\dagger} \hat{a} \rangle$ exhibits substantial system-size-dependent amplification, signaling the emergence of a potential superradiant phase transition as demonstrated in Fig.\ref{fig6}(b) and Fig.\ref{fig6}(c). It is important to note that the photon number distribution of the superradiant state illustrated in Fig.\ref{fig6}(c) only exhibits an even photon number distribution. This suggests that the cavity field in this scenario generates a quadrature-squeezed superradiant state. Notably, our analysis also reveals that macroscopic photon emission persists even under weak coupling conditions, providing robust evidence for the sustainability of superradiant phenomena in this regime. To elucidate the scaling characteristics of $n_{phot}$, we systematically investigated its dependence on varying $k_0$ parameters in the calculations. The observed monotonic enhancement of $n_{phot}$ with increasing $k_0$ unambiguously confirms the manifestation of collective superradiant characteristics. This scaling behavior remains prominently observable in the strong coupling limit, as quantitatively verified in Fig.~\ref{fig6}(d).

	\begin{figure}[htpb]
	\centering
	\includegraphics[width=0.5\textwidth]{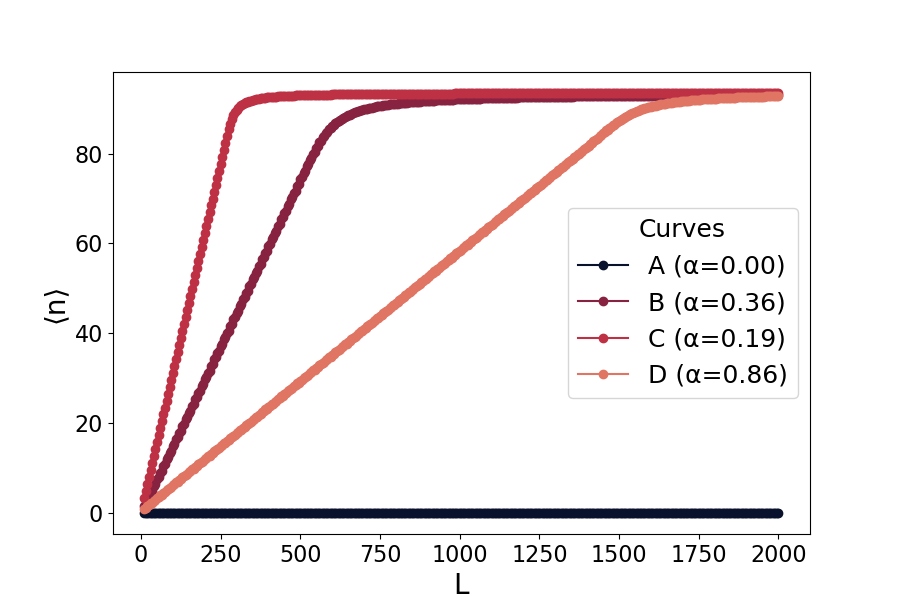} %
	\caption{The average photon number $\langle n\rangle$ of the Bosonic ground state for different system sizes $L$, and the $\alpha$ value fitted to $\langle n\rangle \propto L^{\alpha}$, with other parameters the same as those in Figure \ref{fig4}(b).}
	\label{fig7}
\end{figure}
A detailed investigation of the superradiation behaviors with size effects in Fig.\ref{fig7} demonstrates a clear phase boundary for superradiance emergence and suppression within this framework. Curve A of Fig.\ref{fig7} describes the variation of the cavity photon population (intensity) with system size $L$ when the coupling parameters $k_0$ and $g$ lie outside the superradiant phase transition region. This demonstrates the absence of size-driven superradiance in the non-critical parameter region. Conversely, when the coupling parameters enter the superradiant phase transition regime, the system demonstrates remarkable finite-size effects associated with superradiant transitions. The phase transition characteristics with a power law of $L^\alpha$ observed in curves B, C, and D indicate that the critical system size required for entering the superradiant phase depends on the coupling parameter $g$, where an increasing $g$ reduces the critical size threshold with an increasing power law exponent $\alpha$. The identification of superradiant transitions at subcritical coupling strengths enrich the conventional phase transitions by mixing different order transitions in one system, suggesting the existence of alternative excitation pathways for exploring engineered superradiant states in the solid-state platforms through parameter-controlled symmetry breaking mechanisms, potentially enabling the realization of complex multi-phase transitions in quantum material systems.

\section{ Conclusions and Discussion}

We propose a solvable model in which a tight-binding chain is coupled to a single cavity mode via quantized Peierls substitution. This research uncovers a variety of nonlinear mechanisms for engineering superradiant phase transitions in multicritical solid-state systems. Contrary to the conventional Dicke model, which describes a collection of two-level atoms interacting with a single-mode cavity field and often encounters difficulties in achieving superradiant transitions due to hard handling conditions of light-matter interactions, this model incorporates a momentum-dependent light-matter coupling mechanism. By adjusting the electronic momentum distribution, superradiance can be attained even in the weak coupling regime. Moreover, the non-interacting characteristic of the tight-binding chain enables the acquisition of exact analytical solutions, which clearly elucidate the impact of light-matter coupling on the phase transition and simultaneously avert spurious superradiant phase transitions.

A flexible mechanism for achieving superradiant phase transitions is provided by tuning the center of the electronic momentum distribution (e.g., the momentum symmetric center $k_0$), which obviates the need for precise mode matching and the complex coupling configurations typically required in multi-mode systems \cite{MultimodeOptomechanics}. Experimentally, this single mode cavity setup is more accessible, with superradiant phase transitions observable by simply adjusting system parameters such as coupling strength and momentum distribution. This simplification not only facilitates experimental control but also eliminates the intricate interference effects associated with multimode systems, allowing for the observation of multistability and spontaneous symmetry breaking phenomena\cite{NonMarkovianQuantum}.

Future research in this area has the potential to explore super-radiant phenomena within systems where multimode light fields interact with complex electronic structures that extend beyond their ground states. These investigations may reveal more intricate quantum phases, enhancing our understanding of quantum many-body physics, which is essential for obtaining experimental confirmation. Methodology such as Floquet engineering, which involves the manipulation of the Fermi sea structure through the application of periodic driving fields~\cite{ElectronicFloquetGyro,FloquetDynamicsFermiHubbard}, presents a promising experimental avenue for tunable phase transitions. By adjusting the frequency and amplitude of these driving fields, researchers can dynamically regulate electronic momentum distributions and examine a wide range of phase transitions, thus offering empirical validation for the theoretical perspectives articulated in this study.\\

\section*{ACKNOWLEDGMENTS}

This work was supported by the Natural Science Basic Research Program of Shaanxi province and the National Natural Science Foundation of China for emergency management project (Grants No. 11447025 and No. 11847308).

\end{document}